\documentclass[12pt]{iopart}

 \usepackage{graphicx,times,mathrsfs,amsfonts,cancel}

\begin{document}

\title[Dispersion and damping of 2D dust acoustic waves...]{Dispersion and damping of two-dimensional dust acoustic waves:
Theory and Simulation}

\author{Nitin Upadhyaya}
\address{Department of Applied Mathematics, University of Waterloo, Waterloo, Ontario, Canada N2L 3G1}

\author{Z.\ L.\ Mi\v{s}kovi\'{c}}
\address{Department of Applied Mathematics, University of Waterloo, Waterloo, Ontario, Canada N2L
3G1} \ead{zmiskovi@uwaterloo.ca}

\author{L.-J.\ Hou}
\address{Max-Planck-Institut f\"{u}r Extraterrestrische Physik, 85741 Garching, Germany}

\date{\today}

\begin{abstract}
A two-dimensional generalized hydrodynamics (GH) model is developed to study the full spectrum of both longitudinal and
transverse dust acoustic waves (DAW) in strongly coupled complex (dusty) plasmas, with memory-function-formalism being
implemented to enforce high-frequency sum rules. Results are compared with earlier theories (such as quasi-localized charge
approximation and its extended version) and with a self-consistent Brownian dynamics simulation. It is found that the GH
approach provides good account, not only for dispersion relations, but also for damping rates of the DAW modes in a wide range
of coupling strengths, an issue hitherto not fully addressed for dusty plasmas.
\end{abstract}

\pacs{52.25.Fi, 52.27.Gr, 52.27.Lw}


\maketitle

\section{Introduction}
Dusty, or complex plasmas have grown into a mature research field with a surprisingly broad range of interdisciplinary facets
\cite{Shukla_book,Shukla_2009,Morfill_2009}. There has been significant interest in the study of dynamics
%
of collective excitations of dusty plasmas, both in experiments
\cite{Pieper1996,Homann1997,Nunomura2002,Nunomura2005,Nosenko,Couedel2009,Couedel2010} and through theory/numerical simulation
of strongly-coupled Yukawa systems \cite{Rao_1990,Wang,Kaw,Murillo,Golden_2000,Kalman2000,Ohta2000,
Wang2001,Zhdanov2003,Kalman2004,Piel2006,Hou_2009,Hou2009b}. Of particular interest in these studies is theoretical modeling of
longitudinal and transverse dust acoustic wave (DAW) modes \cite{Rao_1990}, and consequently several assessments of how the
existing analytical theories compare with the results from computer experiments have appeared recently
\cite{Zhdanov2003,Kalman2004,Hou_2009, Hou2009b,Donko_2008}. Specifically, by comparing the Brownian dynamics (BD) simulation
results for the DAW spectra with various theoretical models, it was shown that the so-called Quasi-localized charge
approximation (QLCA) \cite{Golden_2000,Kalman2004} provides an overall good account of the DAW dispersion relations, i.e., the
peak locations in those spectra, for both three-dimensional (3D) and two-dimensional (2D) dusty plasma states characterized by
the coupling strength $\Gamma$ in the range $10<\Gamma<1000$ \cite{Hou_2009,Donko_2008}. However, since the QLCA
provides no information about the profile of such spectra and, consequently, gives no account of damping processes for the
collective modes \cite{Golden_2000}, it is desirable to explore alternate theoretical approaches to strongly coupled systems. In
that context, generalized hydrodynamics (GH) offers possibility to describe full wave spectra by introducing viscoelastic
effects in the dynamics of dusty plasmas in a phenomenological manner, as was demonstrated in studying DAWs in 3D dusty plasmas
\cite{Wang,Kaw,Murillo}.

In contrast to ordinary hydrodynamics (OH), which is only valid for weakly coupled fluids with $\Gamma\ll 1$ in the long
wavelength limit, GH covers wider ranges of coupling strengths and wavelengths, where structural effects begin to show up.
Particularly suitable and physically intuitive framework for implementing the GH model is provided by descriptions of the
Lennard-Jones fluids, pioneered by Ailawadi \textit{et al.}~\cite{Rahman}, Boon and Yip \cite{Boon}, and Hansen and McDonald
\cite{Hansen}. This framework may be applied to a dusty plasma by asserting that dust particles form a quasi-neutral fluid with
the inter-particle interactions described via a Debye-H\"{u}ckel, or Yukawa potential, which arises due to screening of the
charge accumulated on dust particles by the background electron and ion fluids. Successful applications of the GH model to dusty
plasma were conducted by Kaw and Sen \cite{Kaw} and Murillo \cite{Murillo} to study the wave dispersion and the shear wave
cutoff in 3D dust liquids, respectively.

However, there is still great demand for a GH model for 2D strongly coupled dusty plasmas (SCDPs), especially because such
systems have become particularly favored in recent laboratory experiments \cite{Morfill_2009,Piel2006b}.
Therefore, our principal motivation in this work is to develop and implement a GH model for studying the collective dynamics in
2D SCDPs. In addition, it is necessary to generalize the GH model to include the effects of collisions of dust particles with
the neutral gas in the background plasma in a manner similar to that used for colloidal suspensions in describing the collisions
of macroions with the solvent molecules \cite{Hess_1983}. While such collisions give rise to the classical Brownian motion of
macroparticles in both these systems, it should be stressed that the friction forces on dust particles due to the neutrals in
dusty plasmas are generally much weaker than the friction forces on the colloidal macroions. Accordingly, effects of the neutral
gas on the DAW dispersion relations are found to be negligible for the range of parameters of interest in experiments with dusty
plasmas, thus validating the use of molecular dynamics (MD) simulations for such systems \cite{Donko_2008}. However, it is still
unclear as to what extent can damping due to the neutral gas compete with the viscous damping of longitudinal DAWs at finite
frequencies and finite wavelengths in dusty plasmas \cite{Murillo_b}. While this difficult issue naturally arises in this work,
our primary motivation in using BD simulation lies in the fact that treating the dust particles as Brownian particles provides a
natural and convenient way to eliminate the need for thermostation that arises in the MD simulation of dusty plasmas
\cite{Hou_2009,Hou_arXiv}.

In this paper we perform a BD simulation of strongly coupled 2D Yukawa liquids, and use the results to evaluate the equilibrium
radial distribution function (RDF) and the static structure factor, as well as the (power) spectral densities for both the
longitudinal and transverse current densities in such systems. The former two quantities enable us to use the GH approach within
the memory function formalism to evaluate both the dispersion relation and the damping rate of the DAW modes in 2D dusty plasmas
by enforcing the low-order, high-frequency sum rules upon the theoretical spectral densities. It is found that the GH approach
provides dispersion relations that compare well with those resulting from the simulation spectra over broad ranges of
wavelengths and coupling strengths. Additional comparison with the dispersion relations from the QLCA model shows that the GH
approach provides a good account of the direct thermal effect at lower coupling strengths and shorter wavelengths, which is
absent in the QLCA approach, but is seen in the simulation data. On the other hand, a simple extension of the QLCA dispersion
relation, which was recently proposed in Ref.\ \cite{Hou_2009}, is found to be in a much better agreement with both the GH
results and simulation data. Most importantly, the GH results are shown to yield a good fit to the spectral density profiles
from our BD simulations, thus providing semi-analytical modeling of the wave-number dependent damping rates of the DAW modes in
strongly coupled dusty plasmas in a broad range of coupling strengths, an issue that has not been fully addressed so far
\cite{Hou_2009}. Finally, we also find that the effect of collisions with the neutrals on the DAW damping rates is well
accounted for by introducing a local in time friction force into the GH equations.

The manuscript is organized as follows. The details of the BD simulation are given in Section 2. Theoretical development of the
generalized hydrodynamics is discussed in Section 3. Results based on two simple models for the memory function are presented in
Section 4, while we conclude in Section 5.

\section{Simulation}
We perform BD simulation of a 2D system consisting of $N=4000$ dust particles, each carrying a constant charge $q_d$, which are
initially placed at random positions within a square simulation cell in the $\mathbf{r}=(x,y)$ plane with periodic boundary
conditions.
Evolution of the system is governed by equations that may be regarded as a stochastic generalization of the usual MD method,
where the effect of collisions with the neutral particles in the background plasma is modeled by a Langevin generalization of
the usual Newton's equations. Therefore, equations for the velocity and the position vectors of the $i$-th dust particle are
given by
\begin{eqnarray}
\frac{d \mathbf{v}_i(t)}{dt}  &=& -\gamma_n\mathbf{v}_i(t) + \frac{1}{m_d}\sum_{j\neq i}
\mathbf{F}\!\left(\mathbf{r}_i(t)\!-\!\mathbf{r}_j(t)\right) + \mathbf{A}_i(t),
\nonumber \\
\frac{d \mathbf{r}_i(t)}{dt}  &=& \mathbf{v}_i(t), \label{langevin}
\end{eqnarray}
where $m_d$ is the mass of each dust particle and $\mathbf{F}(\mathbf{r})=-(\mathbf{r}/r)dU(r)/dr$ is the pair-wise force
between two dust particles a distance $r=\sqrt{x^2+y^2}$ apart, interacting via the Yukawa potential,
$U(r)=(q_d^2/r)\exp\left(-r/\lambda_D\right)$, with $\lambda_D$ being the Debye screening length of electrons and ions in the
background plasma. Collisions of the $i$-th dust particle with the neutrals in the plasma give rise to a systematic drag force,
which we model by the Epstein drag coefficient $\gamma_n$, and to a cumulative random force, which we model by a
delta-correlated Gaussian white noise. When the system is in thermal equilibrium, the friction coefficient $\gamma_n$ and the
stochastic acceleration of the $i$-th particle $\mathbf{A}_i(t)$ are related to the ambient temperature $T_d$ via the
fluctuation-dissipation theorem. Equations (\ref{langevin}) are solved for our system by the Gear-like predictor-corrector
algorithm for BD simulation \cite{Hou_arXiv}, which was previously used in modeling both the equilibrium structure
\cite{Nitin_2010} and collective modes in 2D dusty plasmas \cite{Hou_2009}.

Equations (\ref{langevin}) can be expressed in terms of just three parameters \cite{Hou_2009}: the coupling strength $\Gamma =
q_d^2/\left(ak_BT_d\right)$, the reduced neutral drag coefficient $\gamma=\gamma_n/\omega_{pd}$, and the screening parameter
$\kappa=a/\lambda_{D}$, where $a\equiv 1/\sqrt{\pi \rho}$ is the average inter-particle separation with $\rho$ being the average
surface density of dust particles, and $\omega_{pd}= \sqrt{2q_d^2/\left(m_da^3\right)}$ is the characteristic dust-plasma
frequency. In this work, we perform BD simulations using $\gamma=0.06$ (which is a value chosen as a matter of convenience only)
and $\kappa$ = 1 as standard values, in a range of coupling strengths $20\le\Gamma\le 1000$ covering liquid-to-crystalline
states of dusty plasma \cite{Hou_2009}.

Further, the current density of dust particles is defined as
\begin{eqnarray}
\mathbf{j}(\mathbf{r},t) = \frac{1}{\sqrt{N}}\sum^{N}_{i=1}\mathbf{v}_i(t)\delta(\mathbf{r}-\mathbf{r}_i(t)),
\end{eqnarray}
with the Fourier transform of its cartesian component $\alpha$ given by
\begin{eqnarray}
j_\alpha(\mathbf{k},t) = \frac{1}{\sqrt{N}}\sum^{N}_{i=1}v_{i\alpha}(t)e^{i\mathbf{k}\cdot\mathbf{r}_i(t)}.
\end{eqnarray}
Assuming that the dust collective modes propagate along the $x$-direction, i.e., $\mathbf{k}=(k,0)$, we let $\alpha=x$ or $y$,
allowing us to define the corresponding longitudinal or transverse current density auto-correlation functions as
\begin{eqnarray}
C_{l,t}(k,t)=\langle j^*_{l,t}(\mathbf{k},0)j_{l,t}(\mathbf{k},t)\rangle, \label{def-correlation}
\end{eqnarray}
where the angular brackets denote ensemble averaging over the initial time. The longitudinal/transverse spectral densities are
then obtained as the Fourier transform of the respective current density auto-correlation functions as,
\begin{eqnarray}
P_{l,t}(k,\omega)=\int^{\infty}_{-\infty}dt\ e^{i\omega t}C_{l,t}(k,t). \label{Power_Spectrum}
\end{eqnarray}
The Fourier transforms defined in Eq.\ (\ref{Power_Spectrum}) are evaluated using Fast Fourier transform from simulation data.
An equivalent and useful definition of the spectral densities, to be used later, is given as the real part of the Laplace
transform \cite{Hansen} of the current density auto-correlation functions,
\begin{eqnarray}
P_{l,t}(k,\omega)=2\Re\{\mathcal{L}[C_{l,t}(k,t)]\}_{s=i\omega}. \label{Laplace_Power}
\end{eqnarray}

\section{Generalized Hydrodynamics}

The basic approach we will take in the development of a GH model for a 2D dusty plasma is to treat the dust system as a
compressible, viscous neutral fluid of particles interacting via the Yukawa potential, analogous to the way how molecules in a
Lennard-Jones fluid interact \cite{Hansen,Murillo_b}. In that context, we note that the OH approach provides a satisfactory
account of the long-time (or low-frequency)
response of a fluid in the weak coupling regime by means of the familiar Naiver-Stokes equation, in which viscous effects are
treated as instantaneous (or local in time) internal friction force, related to fast thermalization of dust particles due to
their mutual collisions. However, as the coupling strength increases, the effect of ``caging'' sets in, so that individual dust
particles are temporarily trapped in potential wells that migrate through the system on a slow time scale, similar to the
picture invoked in the QLCA model. Therefore, the short-time (or high-frequency) response of the system is dominated by elastic
effects due to the restoring forces on dust particles in the itinerant potential wells. In the GH approach to the Naiver-Stokes
equation, a transition from the purely viscous, fluid-like behavior to the elastic response of a solid-like system is described
by postulating a non-local (in time) friction with a memory function characterized by relaxation time $\tau$, such that the
limit of a viscous fluid is recovered at times $t\gg\tau$, whereas the solid-like elastic effects are dominant at times
$t\ll\tau$.

In addition to the collisions among dust particles, it is necessary to include in the GH approach also the effect of their
repeated collisions with the neutral molecules in the background plasma. These collisions may be treated as a Gaussian white
noise, giving rise to a picture of Brownian motion, where each dust particle is subject to a local (in time) friction force with
the neutral drag coefficient $\gamma_n$. We note that for typical dusty plasmas one expects $\gamma_n\tau\ll 1$. On the other
hand, at times $t\ll\gamma_n^{-1}$, the Brownian motion of dust particles due to the neutrals may be considered as ballistic.
Therefore, it is justified to ignore the effect of neutral drag at the time scale $t\sim\tau\ll\gamma_n^{-1}$ where viscoelastic
effects dominate due to interactions among the dust particles \cite{Hess_1983}.
On the other hand, at long times, such that $t\sim\gamma_n^{-1}\gg\tau$, both the viscous drag and the neutral drag on dust
particles may be treated by means of two additive, local in time friction forces in a generalized Navier-Stokes equation because
the underlying physical mechanisms of energy dissipation are statistically independent.

Therefore, we first ignore the neutral drag and introduce a memory function formalism, which incorporates the effects of viscous
relaxation into the short-time dynamics of the dust collective modes by enforcing high-frequency sum rules upon the power
spectral densities of such modes \cite{Boon}. The even-order sum rules are defined in terms of the corresponding frequency
moments as
\numparts
\begin{eqnarray}
\frac{1}{2\pi}\int_{-\infty}^{\infty}d\omega\ \omega^{2n}P_{l,t}(k,\omega) = \langle\omega^{2n}_{l,t}(k)\rangle,
\label{second_sum}
\end{eqnarray}
whereas all odd moments of frequency vanish since $P_{l,t}(k,\omega)$ are even functions of $\omega$. For example, the
zeroth-order sum rule is given by \cite{Boon}
\begin{eqnarray}
\frac{1}{2\pi}\int _{-\infty}^{\infty}d\omega\ P_{l,t}(k,\omega) = v_{th}^2,
\label{first_sum}
\end{eqnarray}
\endnumparts
where $v_{th} = \sqrt{k_BT_d/m_d}$ is the dust thermal speed. A connection with microscopic properties of the system is
accomplished by expressing, e.g., the second moments of frequency for the longitudinal and transverse collective modes in Eq.\
(\ref{second_sum}) with $n=1$ in terms of the RDF of the dust layer, $g(r)$, respectively as \cite{Boon}
\begin{eqnarray}
\langle\omega^2_l(k)\rangle=3k^2v_{th}^4+\frac{\rho}{m_d}v_{th}^2 \int
d^2\mathbf{r}\,g(r)\left[1-\cos\left(kx\right)\right]\frac{\partial^2U(r)}{\partial x^2}, \label{second_moment_RDF}
\end{eqnarray}
\begin{eqnarray}
\langle\omega^2_t(k)\rangle=k^2v_{th}^4+\frac{\rho}{m_d}v_{th}^2 \int
d^2\mathbf{r}\,g(r)\left[1-\cos\left(kx\right)\right]\frac{\partial^2U(r)}{\partial y^2}. \label{second_moment_transverse}
\end{eqnarray}

Once the initial-value properties of the relevant memory functions are fixed by enforcing the sum rules in Eqs.\
(\ref{first_sum}) and (\ref{second_sum}), one may incorporate the effect of the neutral drag by simply adding a local
dissipative force into the fluid equations of motion \cite{Kaw,Murillo_b}, as explained in section \ref{neutral}.

\subsection{Collective Modes}


The starting point for the extension of OH to GH is given by the linearized continuity equation and the Navier-Stokes equation,
which may be written for the longitudinal and transverse dust current densities as \cite{Boon} \numparts
\begin{eqnarray}
\frac{\partial}{\partial t}\delta\!\rho(k,t) - ikj_l(k,t) = 0, \label{k-density}
\end{eqnarray}
\begin{eqnarray}
\frac{\partial}{\partial t}j_l(k,t) - \frac{ik\delta\!\rho(k,t)}{m_d\rho\chi _T} = -\nu _lk^2j_l(k,t),\label{k-current}
\end{eqnarray}
\begin{eqnarray}
\frac{\partial}{\partial t}j_t(k,t) = -\nu _1k^2j_t(k,t), \label{trans-wave}
\end{eqnarray}
\endnumparts
where $\chi_T$ is the isothermal compressibility, $\nu_l = 2\nu _1 + \nu _2$ is defined as the longitudinal viscosity, with
$\nu_1$ being related to the shear viscosity and $\nu_2$ to the bulk viscosity,
$\rho$ is the average surface density, $\delta\!\rho$ is a small perturbation in density, and $j_{l,t}(k,t)$ is a small
perturbation to the longitudinal/transverse current density, assumed to be zero on average.

Inserting the integral of Eq.\ (\ref{k-density}) into Eqs.\ (\ref{k-current}) and (\ref{trans-wave}), one can obtain
integro-differential equations for the current density auto-correlation functions $C_{l,t}(k,t)$ defined in Eq.\
(\ref{def-correlation}), which must be
subject to the initial condition $C_{l,t}(k,0)=v_{th}^2$. In order to model the short-time correlation effects, the resulting
equations can be modified to satisfy the frequency sum rules in the following way
\cite{Rahman,Boon}. In the first step, longitudinal spectral density is forced to satisfy the zeroth frequency sum rule, Eq.\
(\ref{first_sum}), by replacing the isothermal compressibility with the static structure factor, $S(k)$, as
\begin{eqnarray}
\chi_T \mapsto \frac{S(k)}{\rho k_B T_d},
\end{eqnarray}
which generalizes a standard statistical-mechanical relation in the limit $k\to 0$. In the next step, generalizations of the
longitudinal and shear viscosities are introduced via the as yet undefined memory functions, $\phi_l(k,t)$ and $K_t(k,t)$,
respectively,
giving integro-differential equations of the form
\begin{eqnarray}
\frac{\partial}{\partial t}C_{l,t}(k,t) =  -\int _0^tdt'\ K_{l,t}(k,t-t')C_{l,t}(k,t'), \label{new-memory}
\end{eqnarray}
where
\begin{eqnarray}
K_l(k,t) = \frac{(kv_{th})^2}{S(k)} + k^2\phi _l(k,t). \label{memory-kernel}
\end{eqnarray}
It can be shown that the second frequency sum rule, Eq.\ (\ref{second_sum}) with $n=1$, can be satisfied if the initial values
of the longitudinal and transverse viscosity memory functions are expressed in terms of the corresponding second frequency
moments, Eqs.\ (\ref{second_moment_RDF}) and (\ref{second_moment_transverse}),
respectively, as \cite{Boon} \numparts
\begin{eqnarray}
\phi_l(k,0) = \frac{\langle\omega^2_l(k)\rangle}{v_{th}^2k^2} - \frac{v_{th}^2}{S(k)}, \label{initialvalue-memory}
\end{eqnarray}
\begin{eqnarray}
K_t(k,0) = \frac{\langle \omega^2_t(k)\rangle}{v_{th}^2}. \label{trans_init_mem}
\end{eqnarray}
\endnumparts


\subsection{Effects of neutral drag}\label{neutral}

For typical conditions in 2D dusty plasmas, the effect of neutral drag on the dispersion relations of DAW modes is expected to
show up at very low frequencies only, $\omega<\gamma_n$, and we have indeed found in our BD simulations that the actual value of
the drag coefficient $\gamma_n$ has very little effect on the dispersion relations in comparison to the viscoelastic effects.
However, the spectral widths in Eqs.\ (\ref{Power_Spectrum}) or (\ref{Laplace_Power}) are found to be affected by the neutral
drag, as discussed in section \ref{results}.

With the short time dynamics of both the longitudinal and transverse dust collective modes fixed by enforcing the sum rules upon
the initial-value properties of the relevant memory functions, we may now follow suggestions of Refs.\ \cite{Kaw,Murillo_b} and
introduce the long time relaxation effect due to neutral drag
by simply adding a local dissipative term in Eq.\ (\ref{new-memory}), giving
\begin{eqnarray}
\frac{\partial}{\partial t}C_{l,t}(k,t) =  -\int_0^tdt'\ K_{l,t}(k,t-t')C_{l,t}(k,t')-\gamma_n C_{l,t}(k,t).
\label{neutral_drag}
\end{eqnarray}

We proceed with solving Eq.\ (\ref{neutral_drag}) by means of Laplace transform in order to derive expressions for the spectral
density by invoking the definition Eq.\ (\ref{Laplace_Power}). For the longitudinal mode we obtain
\begin{eqnarray}
P_l(k,\omega) =  2v_{th}^2\frac{\omega^2[k^2\phi'_l(k,\omega)+\gamma_n]}{[\omega^2 - \omega _0^2(k) + \omega
k^2\phi''_l(k,\omega)]^2 + \{\omega[k^2\phi'_l(k,\omega)+\gamma_n]\}^2 }, \label{full_spectral_neutral}
\end{eqnarray}
where $\phi'_l(k,\omega)$ and $\phi''_l(k,\omega)$ are the real and imaginary parts, respectively, of the Laplace transformed
longitudinal viscosity memory function,
$\mathcal{L}[\phi_l(k,t)]_{s=i\omega}$,
and where we have explicitly defined
\begin{eqnarray}
\omega_0(k) = \frac{kv_{th}}{\sqrt{S(k)}}. \label{delta_Dispersion}
\end{eqnarray}
Similarly, we obtain for the transverse mode
\begin{eqnarray}
P_t(k,\omega) = 2v^2_{th}\frac{K'_t(k,\omega)+\gamma_n}{[\omega + K''_t(k,\omega)]^2 + [K'_t(k,\omega)+\gamma_n]^2},
\label{trans_power}
\end{eqnarray}
with $K'_t(k,\omega)$ and $K''_t(k,\omega)$ being the real and imaginary parts, respectively, of the Laplace transformed
transverse viscosity memory function, $\mathcal{L}[K_t(k,t)]_{s=i\omega}$.

\subsection{Model Memory Functions}

We note that no approximations were used in the development of the GH approach so far. Specific forms of Eqs.\
(\ref{full_spectral_neutral}) and (\ref{trans_power}) may now be obtained by introducing simple phenomenological models for the
corresponding memory functions.

\subsubsection{Exponential model}

Choosing an exponential longitudinal viscosity memory function of the form $\phi_l(k,t) = \phi_l(k,0)e^{-t/\tau_l}$, we obtain
\cite{Boon}
\numparts
\begin{eqnarray}
\label{exp}
\phi'_l(k,\omega) = \phi_l(k,0)\frac{\tau _l}{1 + \omega^2\tau^2_l},
\end{eqnarray}
\begin{eqnarray}
\label{exp_b}
\phi''_l(k,\omega) = -\phi_l(k,0)\frac{\omega\tau^2_l}{1 + \omega^2\tau^2_l},
\end{eqnarray}
\endnumparts
to be used in Eq.\ (\ref{full_spectral_neutral}). We first discuss the limits of very short and very long viscous relaxation
times $\tau_l$.

In the limit of small relaxation time, $\omega\tau_l\ll 1$, we recover the so-called delta function model for the memory
function be letting $\phi_l(k,0)=\nu_l/\tau_l$, where $\nu_l$ is the longitudinal viscosity. This situation corresponds to a
fluid where viscous drag is described by a local friction force, so that Eq.\ (\ref{full_spectral_neutral}) is reduced to
\cite{Boon}
\begin{eqnarray}
P_l(k,\omega) =  2v_{th}^2\frac{\omega^2\left(k^2\nu _l+\gamma_n\right)}{[\omega^2 - \omega _0^2(k) ]^2 +
\left[\omega\left(k^2\nu _l+\gamma_n\right)\right]^2 }, \label{delta_spectral}
\end{eqnarray}
giving the dispersion relation as $\omega = \omega_0(k)$
with $\omega_0(k)$ defined in Eq.\ (\ref{delta_Dispersion}), and having the full width at half maximum of $\gamma_n+k^2\nu_l$.
This result provides a relatively simple account of the interplay between the neutral drag and viscous damping, assuming that
the quasi-static longitudinal viscosity $\nu_l$ can be defined properly \cite{Donko_2009}.


On the other hand, it is important to study the opposite limit, $\omega\tau_l\gg 1$, for the sake of comparison with the QLCA
model, which is inherently valid in this regime \cite{Kaw_b}. In particular, we find that the viscous damping vanishes in this
limit and the elastic effects in the system's response prevail, giving the dispersion relation $\omega = \omega^l_\infty(k)$
where
\begin{eqnarray}
\omega^l_\infty(k) = \frac{\sqrt{\langle\omega^2_l(k)\rangle}}{v_{th}}, \label{infty_dispersion}
\end{eqnarray}
with the second frequency moment given in terms of the RDF via Eq.\ (\ref{second_moment_RDF}). We note that this situation is
analogous to that in the QLCA model with two important differences: the dispersion relation $\omega = \omega^l_\infty(k)$ is
different from that found in the QLCA model, and the neutral drag still provides a mechanism for damping of the DAW modes.

For finite values of $\tau_l$, a dispersion relation in the exponential model for the longitudinal viscosity memory function may
be obtained from the peak positions in the spectral density given by Eq.\ (\ref{full_spectral_neutral}) with Eqs.\ (\ref{exp})
and (\ref{exp_b}).
This spectral density is fully determined by the functions $S(k)$ and $\langle\omega^2_l(k)\rangle$ via Eqs.\
(\ref{delta_Dispersion}) and (\ref{initialvalue-memory}), respectively, and by the longitudinal viscosity relaxation time
$\tau_l$, which may be a function of $k$ as well. We note that the former two functions can be calculated, at least in
principle, from the RDF of the dust layer by using the usual definition for $S(k)$, and Eq.\ (\ref{second_moment_RDF}) for
$\langle\omega^2_l(k)\rangle$.
However, while $g(r)$ can be obtained directly from BD simulations, or may be available from first principles, there is no
simple way to determine $\tau_l$ from first principles. One possibility to proceed is to ignore its $k$ dependence and treat
$\tau_l$ as a free parameter, possibly dependent on $\Gamma$, that can be determined from, e.g., fitting the peak positions of
Eq.\ (\ref{full_spectral_neutral}) to the peak positions of the spectral densities obtained from experiments or computer
simulations. It is expected that such a procedure would yield a dispersion relation lying somewhere between $\omega =
\omega_0(k)$ and $\omega = \omega^l_\infty(k)$, defined via Eqs.\ (\ref{delta_Dispersion}) and (\ref{infty_dispersion}),
respectively.

Implementation of the exponential model for transverse viscosity memory function  $K_t(k,t)$ is a straightforward repetition of
the procedure outlined above, and it produces a result for the spectral density that is entirely equivalent to that derived by
Murillo for transverse modes in 3D strongly coupled Yukawa liquids \cite{Murillo}. In brief, one assumes $K_t(k,t) =
K_t(k,0)e^{-t/\tau_t}$ with the initial value given in Eq.\ (\ref{trans_init_mem}), where the second moment of frequency for the
transverse mode may also be obtained from the RDF by using Eq.\ (\ref{second_moment_transverse}).
One further obtains the real and imaginary parts of the Laplace transform $\mathcal{L}[K_t(k,t)]_{s=i\omega}$
by using expressions analogous to Eqs.\ (\ref{exp}) and (\ref{exp_b}), which, upon substitution into Eq.\ (\ref{trans_power})
yield a spectral density for the transverse mode in the exponential model.

It should be reiterated that, in the case of vanishing relaxation time, $\tau_t=0$, the transverse mode is purely diffusive and
the resulting spectral density in Eq.\ (\ref{trans_power}) exhibits no dispersion \cite{Boon}.
In the opposite limit of $\tau_t\to \infty$, viscous damping vanishes, while the dispersion relation, given by $\omega =
\omega^t_\infty(k)$ with
\begin{eqnarray}
\omega^t_\infty(k)=\frac{\sqrt{\langle\omega^2_t(k)\rangle}}{v_{th}}, \label{infty_dispersion_transverse}
\end{eqnarray}
can be shown to exhibit a quasi-acoustic behavior in the long wavelength limit. On the other hand, when using the peak positions
of the spectral density in Eq.\ (\ref{trans_power}) for the exponential model with finite $\tau_t$, one encounters a cutoff
wavenumber, $k_c$, in the dispersion relation for the transverse mode, which can be estimated from the condition
$\langle\omega^2_t(k_c)\rangle=\left(v_{th}/\tau_t\right)^2$ in the limit of vanishing neutral drag \cite{Murillo,Hou_2009}.
However, like in the case of the longitudinal mode, since there is no simple way to determine $\tau_t$ from first principles,
one can again treat it as a free parameter that may be determined from an appropriate fitting, e.g., by using the cutoff values
$k_c$ observed in the experiments on shear waves in 2D Yukawa liquids \cite{Nosenko}.

\subsubsection{Gaussian model}

Instead of using $\tau_l$ and $\tau_t$ as free parameters, one may attempt enforcing higher-order frequency sum rules to see if
a refinement of the above exponential model can be achieved. However, since the exponential model for the viscosity memory
functions does not support moments higher than the second order when used in Eqs.\ (\ref{full_spectral_neutral}) or
(\ref{trans_power}), one needs a model with different time dependence. Besides having to satisfy the second-frequency sum rule
via Eq.\ (\ref{initialvalue-memory}), it is shown in the Appendix that such model must additionally satisfy both the third-order
sum rule giving $\left[\frac{\partial}{\partial t}\phi_l(k,t)\right]_{t=0}=0$, and the fourth-order sum rule giving
\begin{eqnarray}
\left[\frac{\partial^2}{\partial t^2}\phi _l(k,t)\right]_{t=0}  =
\frac{\frac{\langle\omega^2_l(k)\rangle^2}{v^2_{th}}-\langle\omega^4_l(k)\rangle}{k^2v^2_{th}}.  \label{memory_2nd_initial}
\end{eqnarray}

We see now that a Gaussian model for the longitudinal viscosity memory function of the form
$\phi_l(k,t)=\phi_l(k,0)\exp\left[-t^2/\sigma^2_l(k)\right]$ will satisfy all moments up to and including the fourth if the
Gaussian relaxation time $\sigma_l(k)$ is given by
\begin{eqnarray}
\sigma^2_l(k)=\frac{2k^2v^4_{th}\phi_l(k,0)}{v^2_{th}\langle\omega^4_l(k)\rangle-\langle\omega^2_l(k)\rangle^2},\label{Gauss_time}
\end{eqnarray}
with the initial value of the memory function, $\phi_l(k,0)$, given by Eq.\ (\ref{initialvalue-memory}).
Finally, by using the Laplace transform of the Gaussian longitudinal viscosity memory function with $s=i\omega$ in Eq.\
(\ref{full_spectral_neutral}), we obtain a spectral density which is fully determined by three functions: $S(k)$,
$\langle\omega^2_l(k)\rangle$, and $\langle\omega^4_l(k)\rangle$ via Eqs.\ (\ref{delta_Dispersion}) and
(\ref{initialvalue-memory}) without the need for free parameters. Unfortunately, analytical expressions that can be used for
calculation of the fourth frequency moment are rather cumbersome and require a three-particle distribution function, which is
difficult to obtain from simulations \cite{Boon}. Therefore, since calculation of $\langle\omega^4_l(k)\rangle$ from first
principles is impractical, we may use the Gaussian model by computing the fourth moment numerically from Eq.\ (\ref{second_sum})
with $n=2$
where $P_l(k,\omega)$ is obtained from simulation. For consistency, it is then desirable to also compute the second moment from
Eq.\ (\ref{second_sum}) with $n=1$ using the same $P_l(k,\omega)$ based on simulation data. In this way, we may use the Gaussian
model as a simulation-based, parameter-free test of the quality of the approximation achieved in using the exponential model
with a suitable choice of finite relaxation time $\tau_l$.

We finally note that a completely analogous development of the Gaussian model can be applied to the transverse mode.

\section{Results and Discussion}
\label{results}

We tested several values for the reduced neutral drag coefficient $\gamma$ in our BD simulation and found no noticeable
dependence of the resulting spectra on $\gamma$. Moreover, all the quantities used in the GH modeling of the collective modes
(e.g., the RDF and the relaxation times in the exponential model of the memory functions for the longitudinal and transverse
modes) were also found to be robustly independent of the (small) values of $\gamma$ used in simulation.
Therefore, all results will be shown for a standard value of $\gamma=0.06$, along with the standard screening parameter
$\kappa=1$.

\subsection{Longitudinal Wave Mode}

We use coupling strength with values $\Gamma=$ 20, 60, 100, 200, 600, and 1000 to investigate the longitudinal mode in a broad
range of dusty plasma conditions, going from liquid to crystalline states.

\begin{figure}
\begin{center}
\includegraphics[width=\textwidth]{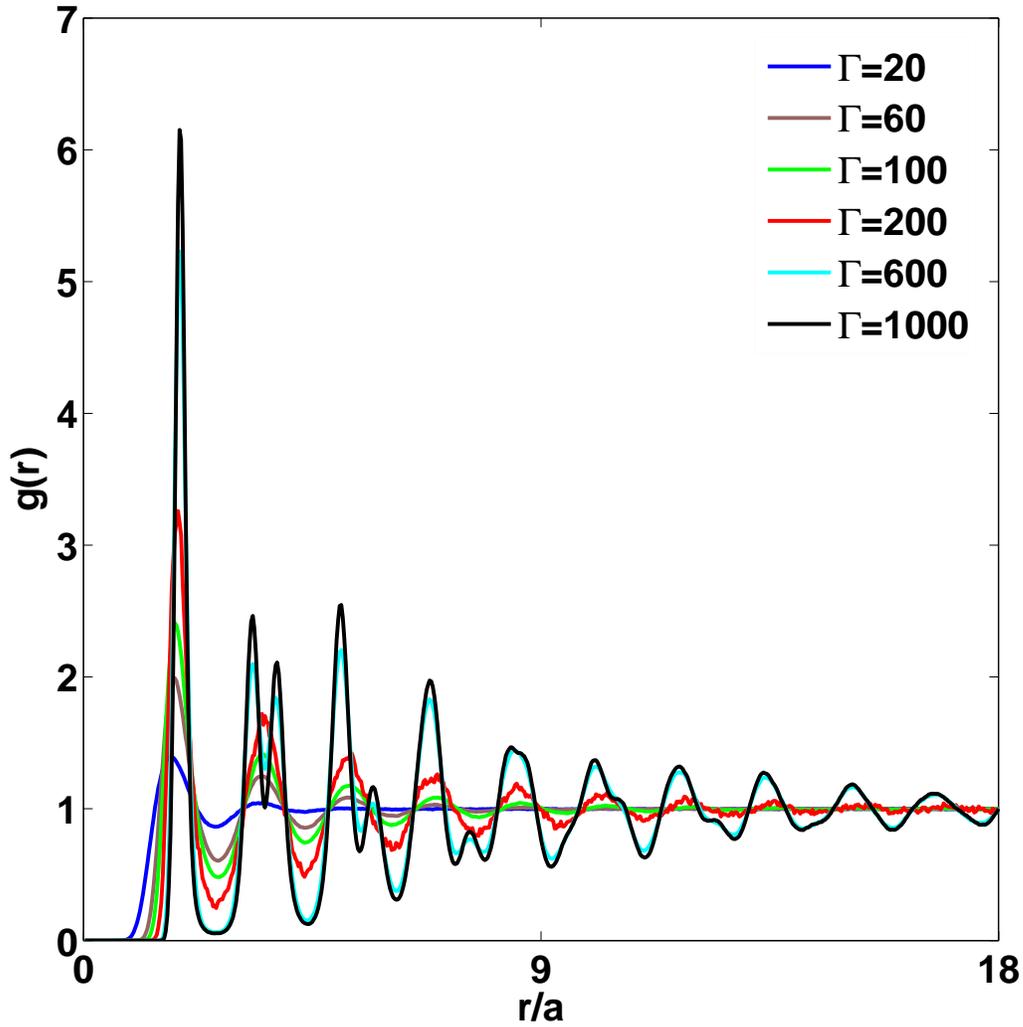}
\caption{\small \label{fig:gr} Radial distribution function $g(r)$ as a function of the reduced distance $r/a$ for $\kappa=1.0$,
$\gamma=0.06$, and $\Gamma=20$, 60, 100, 200, 600, and 1000, corresponding to the increasing hight of the principal peak,
respectively.}
\end{center}
\end{figure}

\begin{figure}
\begin{center}
\includegraphics[width=\textwidth]{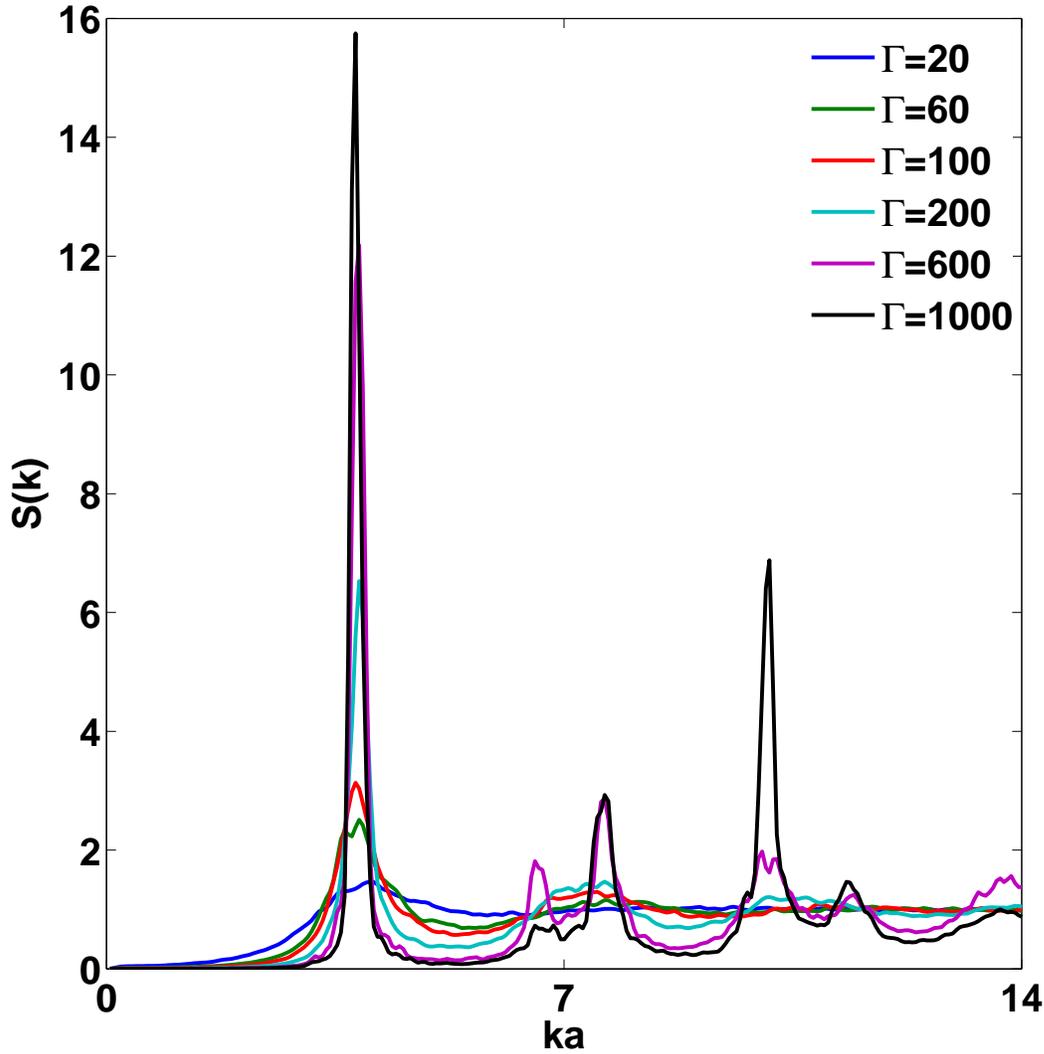}
\caption{\small \label{fig:structure_Factor}Static structure factor $S(k)$ as a function of the reduced wave-number $ka$ for
$\kappa=1.0$, $\gamma=0.06$, and $\Gamma=20$, 60, 100, 200, 600, and 1000, corresponding to the increasing hight of the
principal peak, respectively.}
\end{center}
\end{figure}

As described above, both the static structure factor $S(k)$ and the second frequency moments can be calculated from the
equilibrium RDF $g(r)$, which is shown in Fig.\ \ref{fig:gr} for several values of $\Gamma$. However, there are significant
difficulties related to using $S(k)$ for modeling the dispersion relation of the longitudinal mode at long wavelengths. Namely,
when $S(k)$ is calculated from the RDF, its small-$k$ values exhibit a very strong dependence on the fluctuations appearing in
the simulation due to the finite number of dust particles \cite{Hartmann}. This is expected to give rise to rather noisy
dispersion curves, especially in the situations characterized by short relaxation times when the dispersion is dominated by the
frequency $\omega_0(k)$ defined in Eq.\ (\ref{delta_Dispersion}). To remedy the situation to some extent, we resort to
calculating $S(k)$ directly from the simulation data by using the long time average of the Fourier transformed particle density
in equilibrium, as follows
\begin{eqnarray}
S(k)=\langle \rho^*(\mathbf{k})\rho(\mathbf{k})\rangle, \label{structure}
\end{eqnarray}
where
\begin{eqnarray}
\rho(\mathbf{k}) = \frac{1}{\sqrt{N}}\sum^{N}_{i=1}e^{i\mathbf{k}\cdot\mathbf{r}_i}.
\end{eqnarray}
In Fig.\ \ref{fig:structure_Factor}, we show the thus obtained results for $S(k)$ for several $\Gamma$ values, and note that the
peak structures in that figure were found to be identical to those arising in $S(k)$ computed from the corresponding RDFs in
Fig.\ \ref{fig:gr}.

\begin{figure}
\begin{center}
\includegraphics[width=\textwidth]{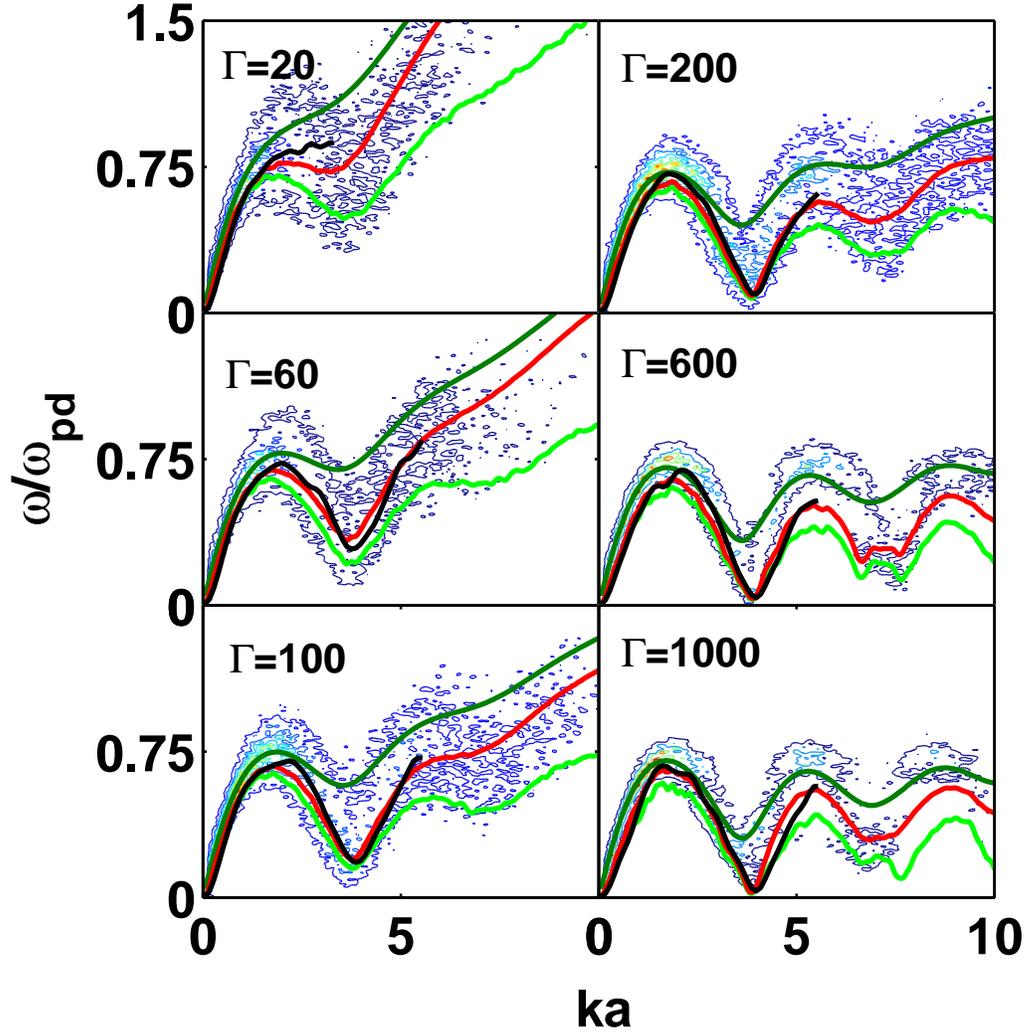}
\caption{\small \label{fig:Dispersion_1} Longitudinal wave dispersion curves against simulation spectra for $\kappa=1.0$,
$\gamma=0.06$, and $\Gamma=20$, 60, 100, 200, 600, and 1000. The upper (dark green), middle (red) and the lower (light green)
solid curves represent the exponential model with an infinitely long, finite, and vanishing relaxation times, respectively. The
thick solid (black) curve represents the Gaussian model for wave-numbers $ka < 6$. }
\end{center}
\end{figure}

In Fig.\ \ref{fig:Dispersion_1} we show the simulation data for the spectral density of the longitudinal mode, along with the
dispersion curves from the exponential model, represented by three solid curves corresponding to, in descending order,
infinitely long, finite, and vanishing relaxation times $\tau_l$.
Also shown in Fig.\ \ref{fig:Dispersion_1} by a thick black solid line is the dispersion curve obtained from the Gaussian model
with the second and fourth frequency moments evaluated from Eq.\ (\ref{second_sum}) with $n=1$ and 2, respectively, where
$P_l(k,\omega)$ was taken directly from the simulation data for spectral density. While the three curves for the exponential
model extend over the full range of wave-numbers shown in Fig.\ \ref{fig:Dispersion_1}, the results for the Gaussian model are
shown over a reduced range of wave-numbers covering the first Brillouin zone only, $ka< 6$, because computations of the
frequency moments via Eq.\ (\ref{second_sum}) were hampered by a significant noise in the simulation spectra at larger
wave-numbers and lower coupling strengths.


One notices in Fig.\ \ref{fig:Dispersion_1} that, for the exponential model with vanishing relaxation time (or, equivalently,
the delta-function model), the resulting dispersion relation, $\omega=\omega_0(k)$ with $\omega_0(k)$ given in Eq.\
(\ref{delta_Dispersion}), displays some noise coming from the simulation data via Eq.\ (\ref{structure}). We found this noise to
be significantly weaker than the noise arising when $S(k)$ is calculated from the RDF. On the other hand, as the longitudinal
relaxation time $\tau_l$ increases in the exponential model, there is a relative increase in the contribution of the second
frequency moment $\langle\omega^2_l(k)\rangle$, which, when evaluated from the RDF via Eq.\ (\ref{second_moment_RDF}), exhibits
a much smoother $k$ dependence than $\omega_0(k)$ with $S(k)$ evaluated from Eq.\ (\ref{structure}).

As mentioned before, the dispersion relation for finite values of the relaxation time $\tau_l$ in the exponential model for the
longitudinal viscosity memory function is obtained from the peak positions in the spectral density given by Eq.\
(\ref{full_spectral_neutral}) with Eqs.\ (\ref{exp}), (\ref{exp_b}), (\ref{delta_Dispersion}), and (\ref{initialvalue-memory}).
We have chosen the values of $\tau_l$ which provide the best fit to the peaks in the simulation spectral densities, and we have
found $\tau_l$ to be a relatively weak function of $\Gamma$ that may be reasonably well approximated by
\begin{eqnarray}
\tau_l(\Gamma) = \cases{
  0.769 & for $\Gamma\le 60$ \\
  1.000 & for $60<\Gamma\le 200$ \\
  1.333 & for $200<\Gamma\le 1000$. \\}
\label{tau_l}
\end{eqnarray}
We emphasize that the quality of our choice of the values for $\tau_l$ in the exponential model is confirmed through a close
agreement with the dispersion curves from the Gaussian model for wave-numbers in the first Brillouin zone, as displayed in Fig.\
\ref{fig:Dispersion_1}.

\begin{figure}
\begin{center}
\includegraphics[width=\textwidth]{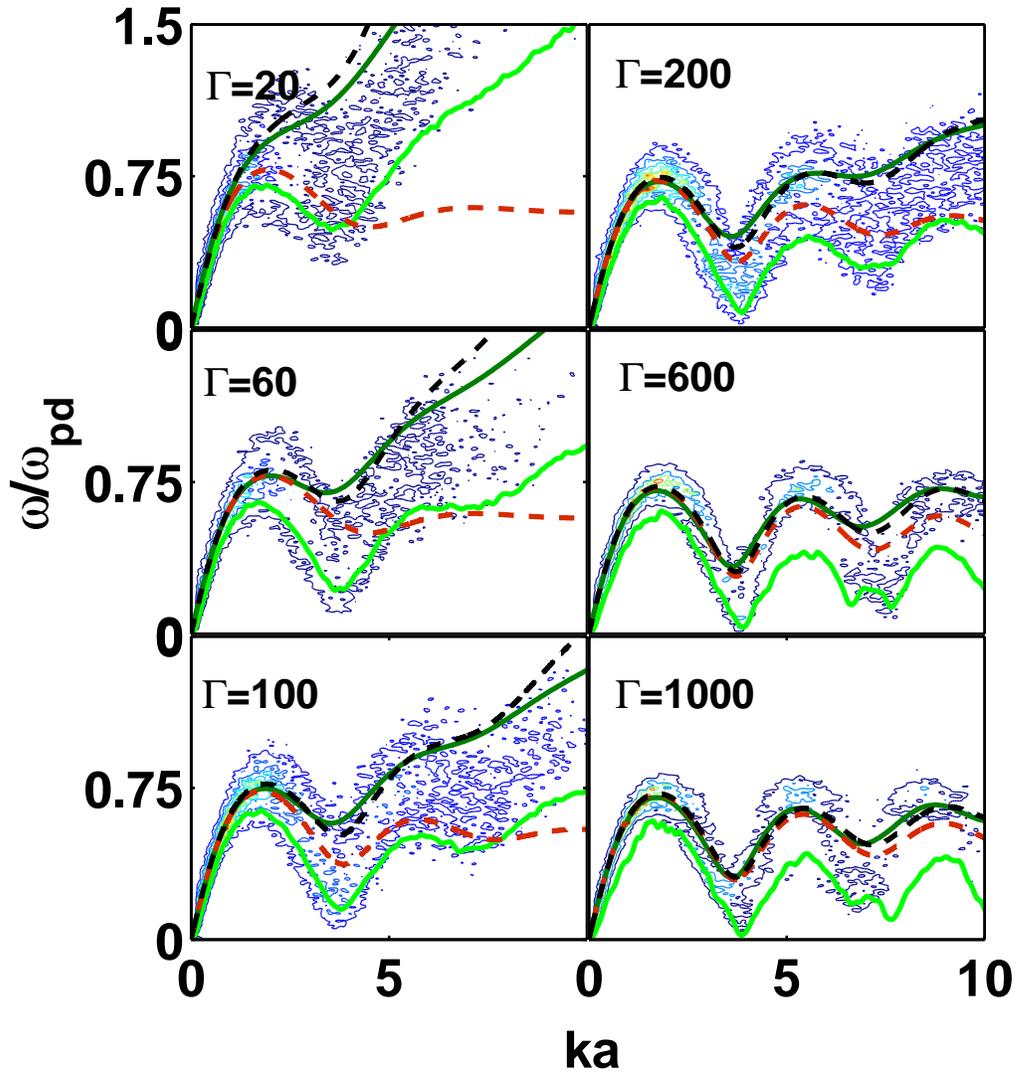}
\caption{\small \label{fig:Dispersion_2} Longitudinal wave dispersion curves against simulation spectra for $\kappa=1.0$,
$\gamma=0.06$, and $\Gamma=20$, 60, 100, 200, 600, and 1000. The upper (black) and the lower (red) dashed curve represent the
results from the EQLCA and QLCA models, respectively. The upper (dark green) and the lower (light green) solid curves represent
the exponential model with an infinitely long and vanishing relaxation times, respectively.}
\end{center}
\end{figure}

It is worth mentioning that the extreme cases of the exponential model, corresponding to the limits $\tau_l\to 0$ and
$\tau_l\to\infty$, yield two parameter-free dispersion relations for the longitudinal mode, $\omega=\omega_0(k)$ and
$\omega=\omega^l_\infty(k)$, respectively. Remarkably, it is noticed in Fig.\ \ref{fig:Dispersion_1} that these two dispersion
relations provide good account of the simulation data
by straddling the thermal noise, seen in the recorded spectra at the low-to-medium $\Gamma$ values. Moreover, at the
medium-to-high $\Gamma$ values, one notices that $\omega=\omega_0(k)$ closely follows the dispersion relation from the Gaussian
model for the wave-numbers in the first Brillouin zone and, in particular, reproduces the near-vanishing of frequency at around
$ka\approx 4$ for $\Gamma=600$ and 1000. This is somewhat surprising, given that the delta-function model is stretched well into
the condensed state at those two coupling strengths.

We further compare in Fig.\ \ref{fig:Dispersion_2} the dispersion relations $\omega=\omega_0(k)$ and $\omega=\omega^l_\infty(k)$
(shown by the upper and lower solid curves, respectively) with the results from the QLCA model \cite{Donko_2008}, shown by the
lower dashed curve. It is immediately obvious that the QLCA dispersion does not reproduce the direct thermal effect, which is
responsible for a quasi-linear increase in the peak frequencies of the simulation spectra at higher wave-numbers and lower
coupling strengths. In a previous work, we have shown that this deficiency can be easily rectified by a simple extension of the
QLCA model, labeled as the EQLCA model in Ref.\cite{Hou_2009}, which is shown by the upper dashed curve in Fig.\
\ref{fig:Dispersion_2}.
One notices a surprisingly good agreement between the EQLCA dispersion relation and the curve $\omega=\omega^l_\infty(k)$ from
the exponential model for $\tau_l\to\infty$. While an agreement with the QLCA dispersion relation was expected at the higher
$\Gamma$ values and lower $k$ values owing to the fact that the QLCA model inherently assumes $\omega\tau_l\gg1$ \cite{Kaw_b}
and the direct thermal effect is relatively weak at low $k$ and high $\Gamma$ values \cite{Hou_2009}, it is remarkable how the
parameter-free dispersion relation $\omega=\omega^l_\infty(k)$ provides justification for the empirically derived EQLCA model
over the broad ranges of wavenumbers and coupling strengths.

\begin{figure}
\centering
\includegraphics[width=\textwidth]{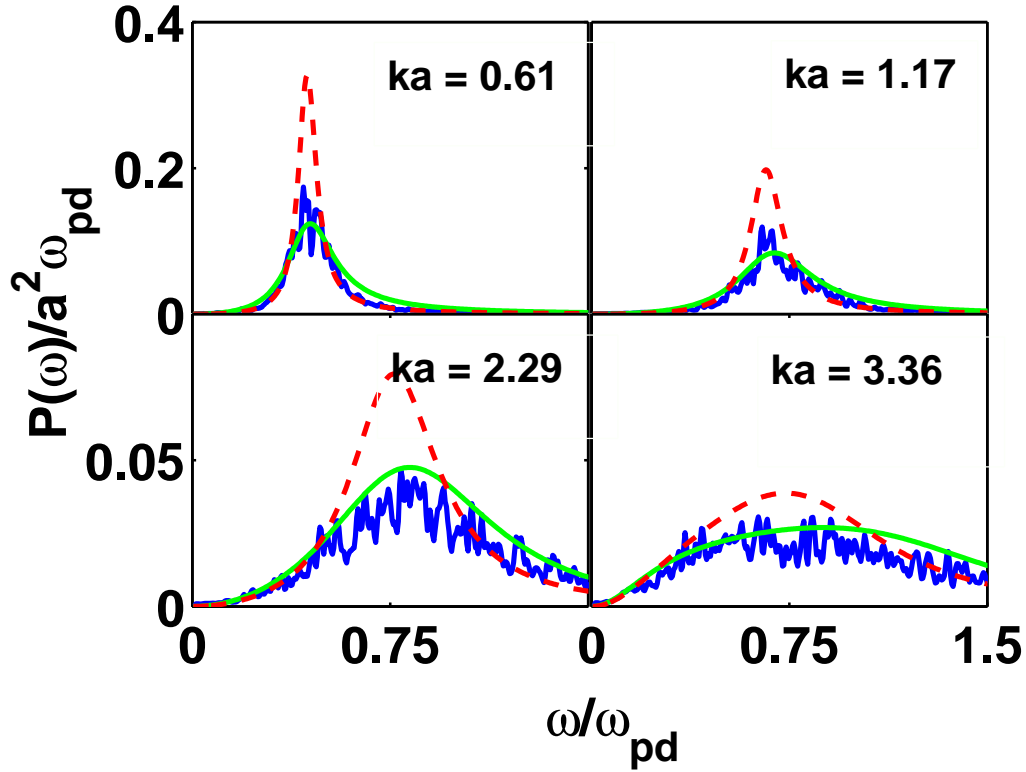}
\caption{ \small \label{fig:profileCurves_20} Longitudinal spectral density profile curves versus reduced frequency
$\omega/\omega_{pd}$ for $\kappa=1.0$, $\gamma=0.06$, $\Gamma=20$, and $ka=0.61, 1.17, 2.29$, and 3.36. Simulation data are
shown by the noisy (blue) solid curve, exponential model with finite relaxation time is represented by the dashed (red) curve,
and the Gaussian model is represented by the smooth solid (green) curve. }
\end{figure}

\begin{figure}
\begin{center}
\includegraphics[width=\textwidth]{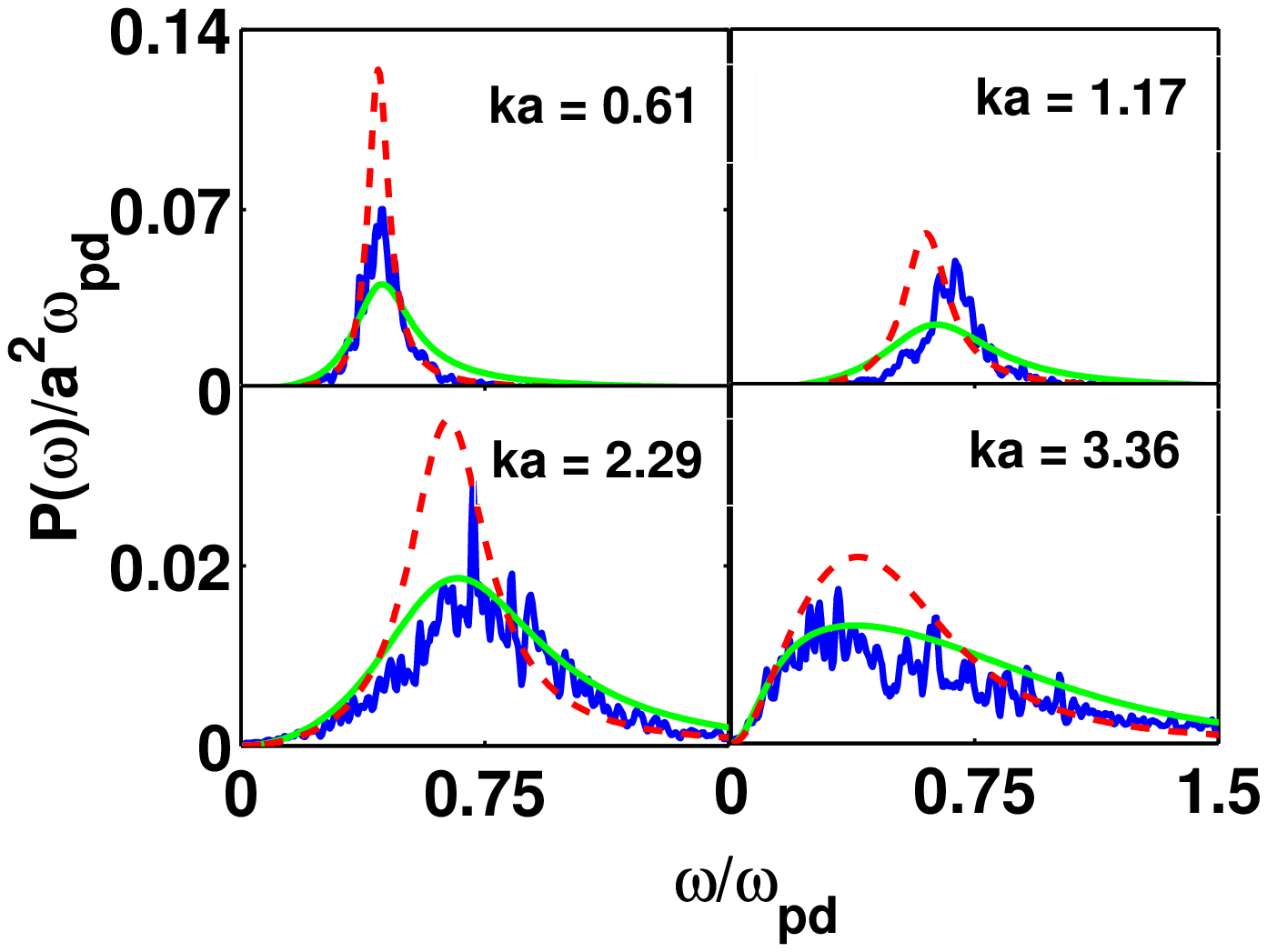}
\caption{\small \label{fig:profileCurves_60} The same as Fig.\ \ref{fig:profileCurves_20}, but for $\Gamma=60$.}
\end{center}
\end{figure}

\begin{figure}
\begin{center}
\includegraphics[width=\textwidth]{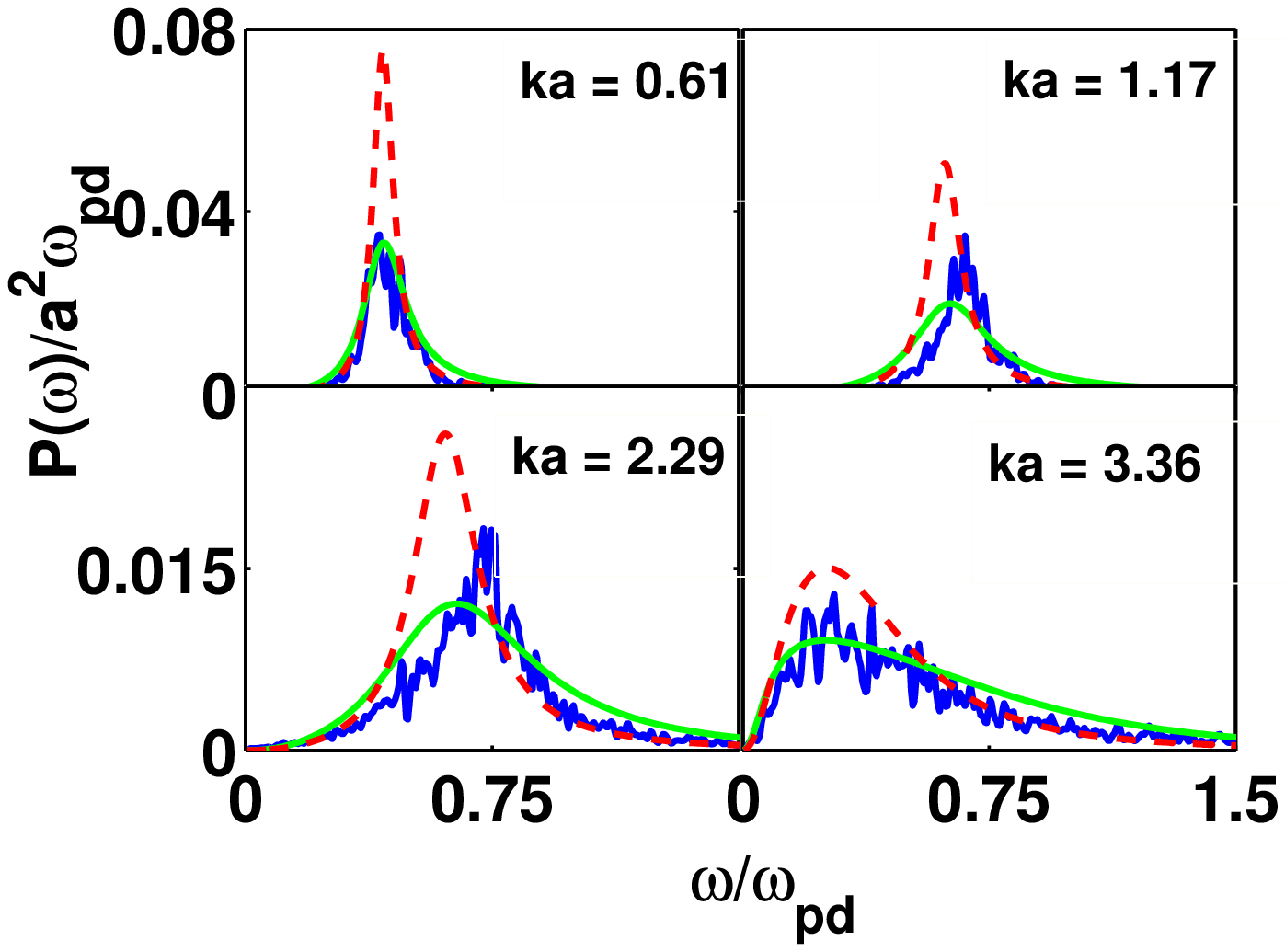}
\caption{\small \label{fig:profileCurves_100} The same as Fig.\ \ref{fig:profileCurves_20}, but for $\Gamma=100$.}
\end{center}
\end{figure}

\begin{figure}
\begin{center}
\includegraphics[width=\textwidth]{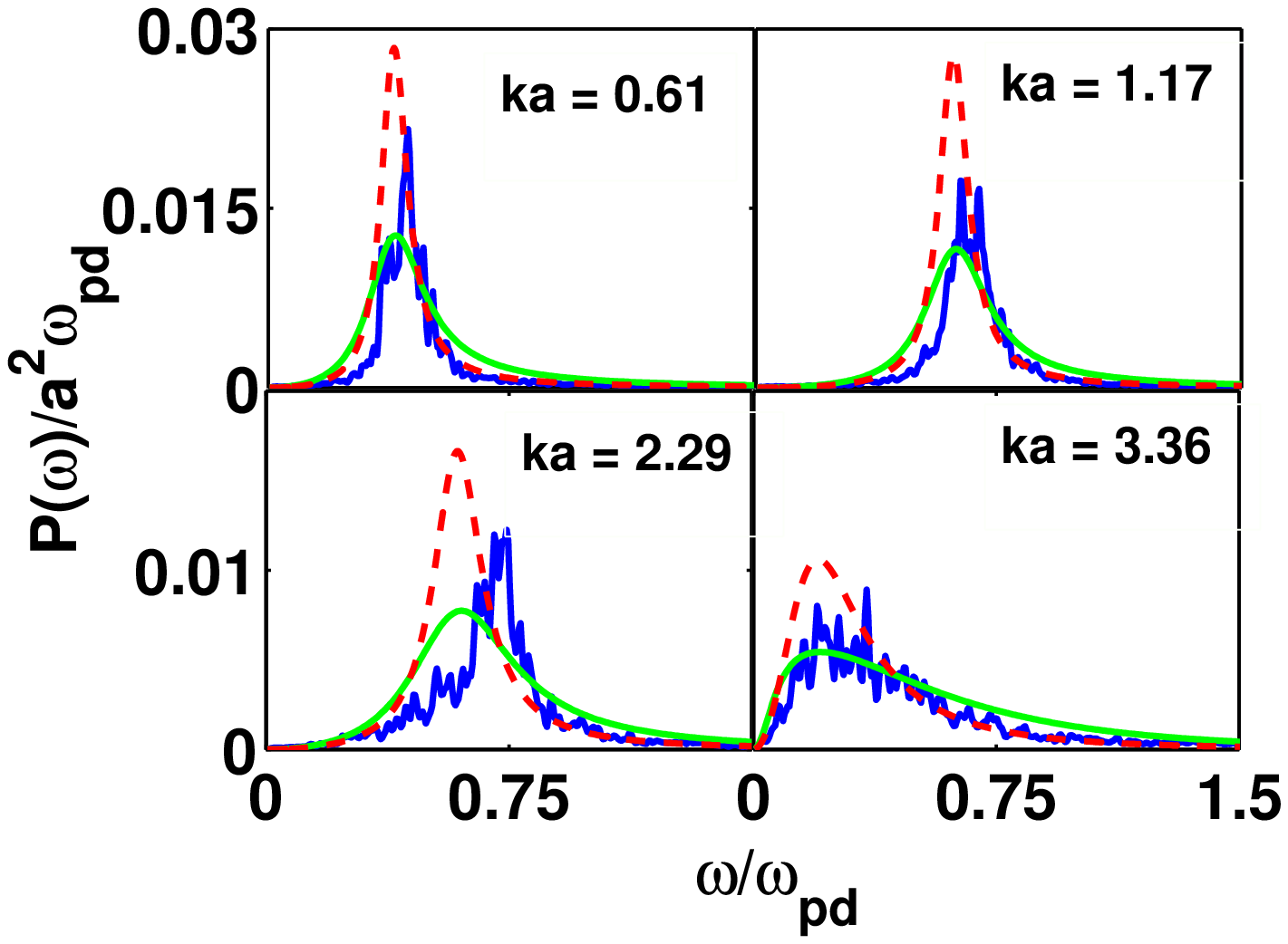}
\caption{\small \label{fig:profileCurves_200} The same as Fig.\ \ref{fig:profileCurves_20}, but for $\Gamma=200$.}
\end{center}
\end{figure}

\begin{figure}
\begin{center}
\includegraphics[width=\textwidth]{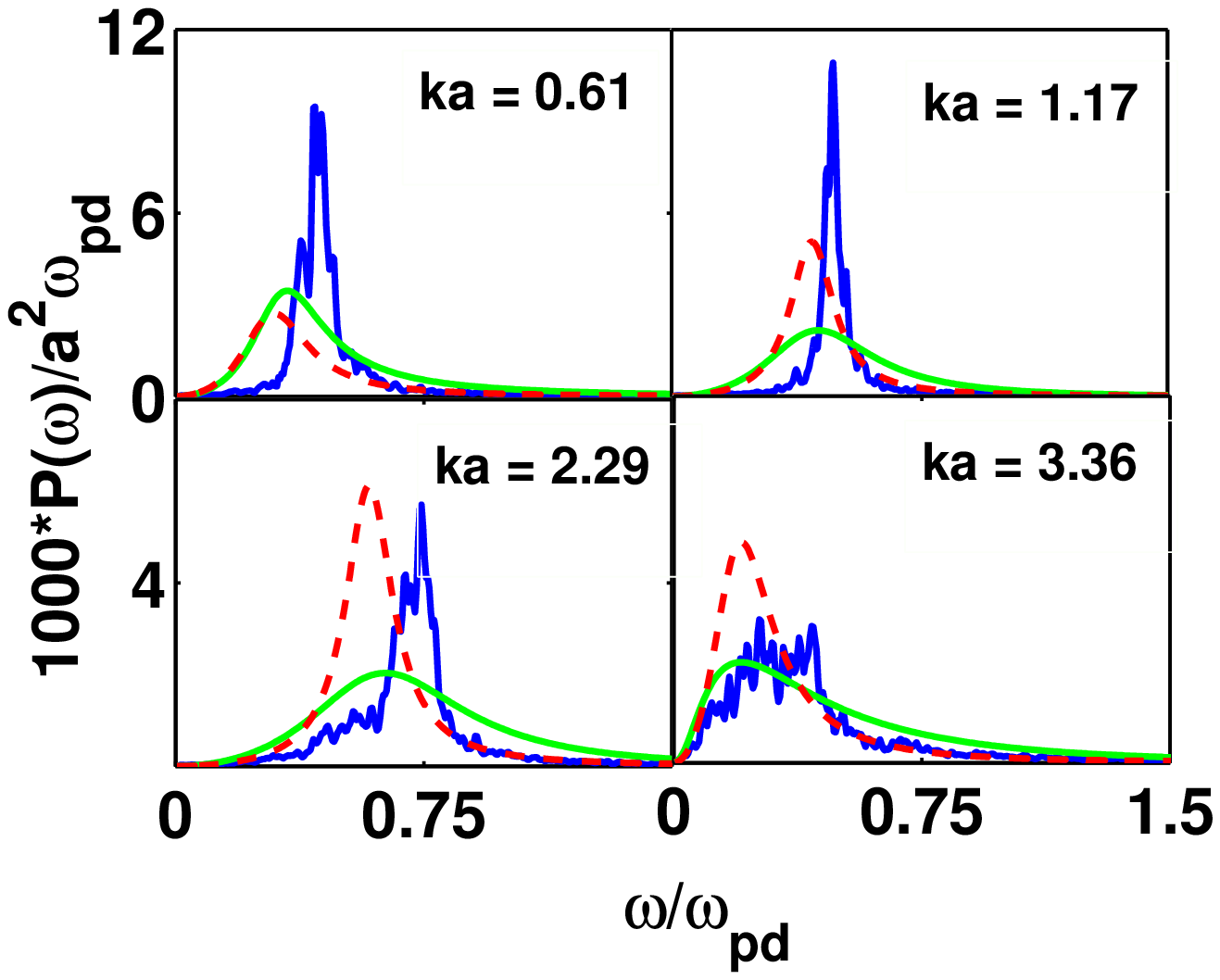}
\caption{\small \label{fig:profileCurves_600} The same as Fig.\ \ref{fig:profileCurves_20}, but for $\Gamma=600$.}
\end{center}
\end{figure}

\begin{figure}
\begin{center}
\includegraphics[width=\textwidth]{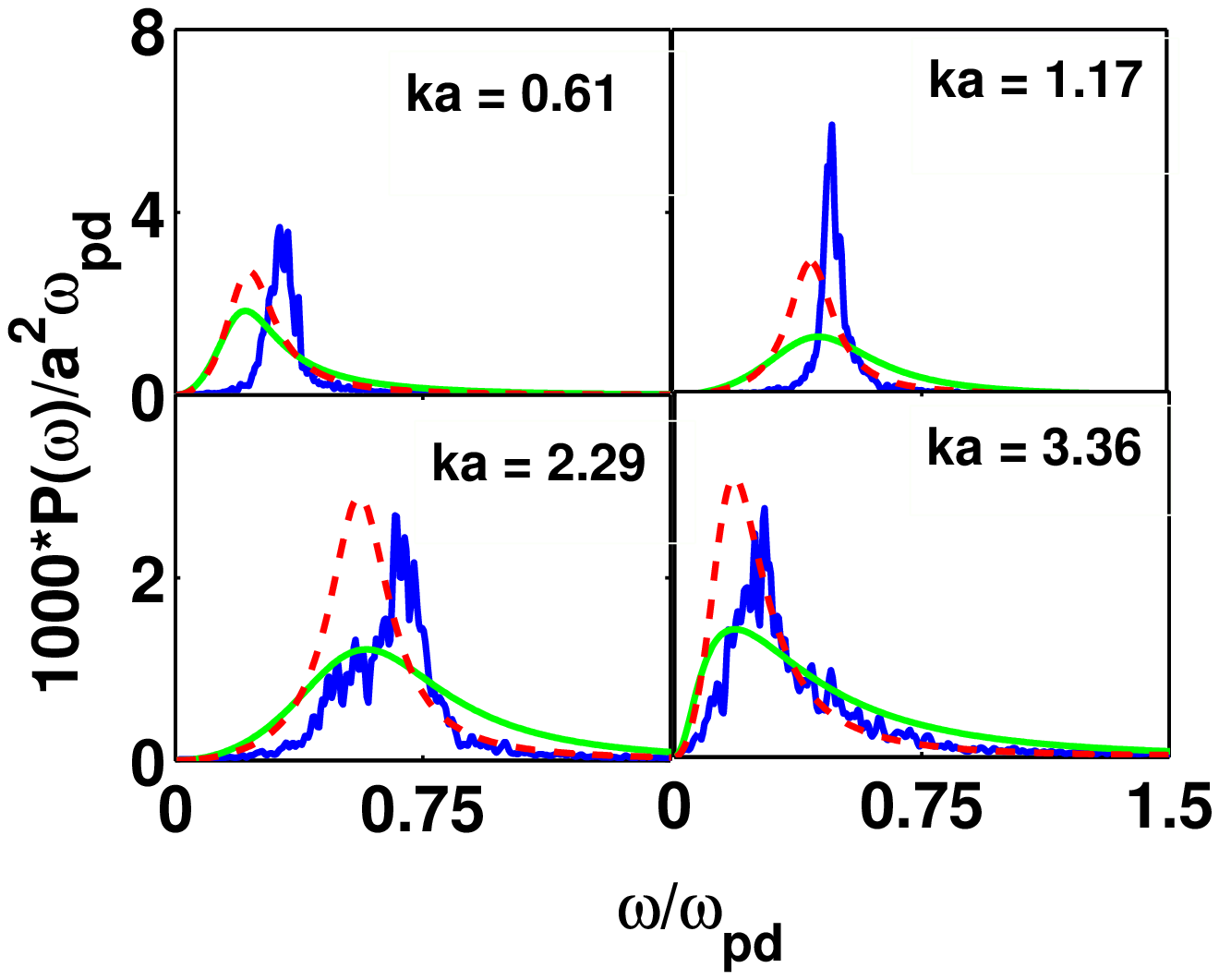}
\caption{\small \label{fig:profileCurves_1000} The same as Fig.\ \ref{fig:profileCurves_20}, but for $\Gamma=1000$.}
\end{center}
\end{figure}

We next compare the performances of the exponential model with finite relaxation time $\tau_l$ from Eq.\ (\ref{tau_l}) and the
Gaussian model with the frequency moments evaluated from the simulation spectra via Eq.\ (\ref{second_sum}) by using these two
models in Eq.\ (\ref{full_spectral_neutral}) to predict the frequency dependent spectral profiles at several fixed wave-numbers.
Given that the half-width at half-maximum (HWHM) of these profiles may be directly related to the wave-number dependent damping
rate of the longitudinal DAW mode, in this way we demonstrate the advantage of using the memory function formalism within the GH
model in tackling the difficult issue of damping of the collective excitation modes in dusty plasmas.

Specifically, we show in Figs.\ 5-10 the frequency dependencies obtained from the spectral density in Eq.\
(\ref{full_spectral_neutral}) with the exponential and the Gaussian models for four wave-numbers, $ka=0.61, 1.17, 2.29$, and
3.36, and compare them with the corresponding profiles of the simulation spectral density
for $\Gamma=$ 20, 60, 100, 200, 600, and 1000, respectively. One notices in Figs.\ 5-10 a fair agreement between the simulation
data and both GH models, which is especially good for lower $\Gamma$ values where the hydrodynamic regime is expected to be more
pronounced, while for very high coupling strengths, say, $\Gamma> 200$, the agreement begins to deteriorate. Comparison of both
GH models with the simulation data may be considered fairly reasonable, even at high coupling strengths where the hydrodynamic
model has been stretched well into the crystalline state.

\begin{figure}
\centering
\includegraphics[width=\textwidth]{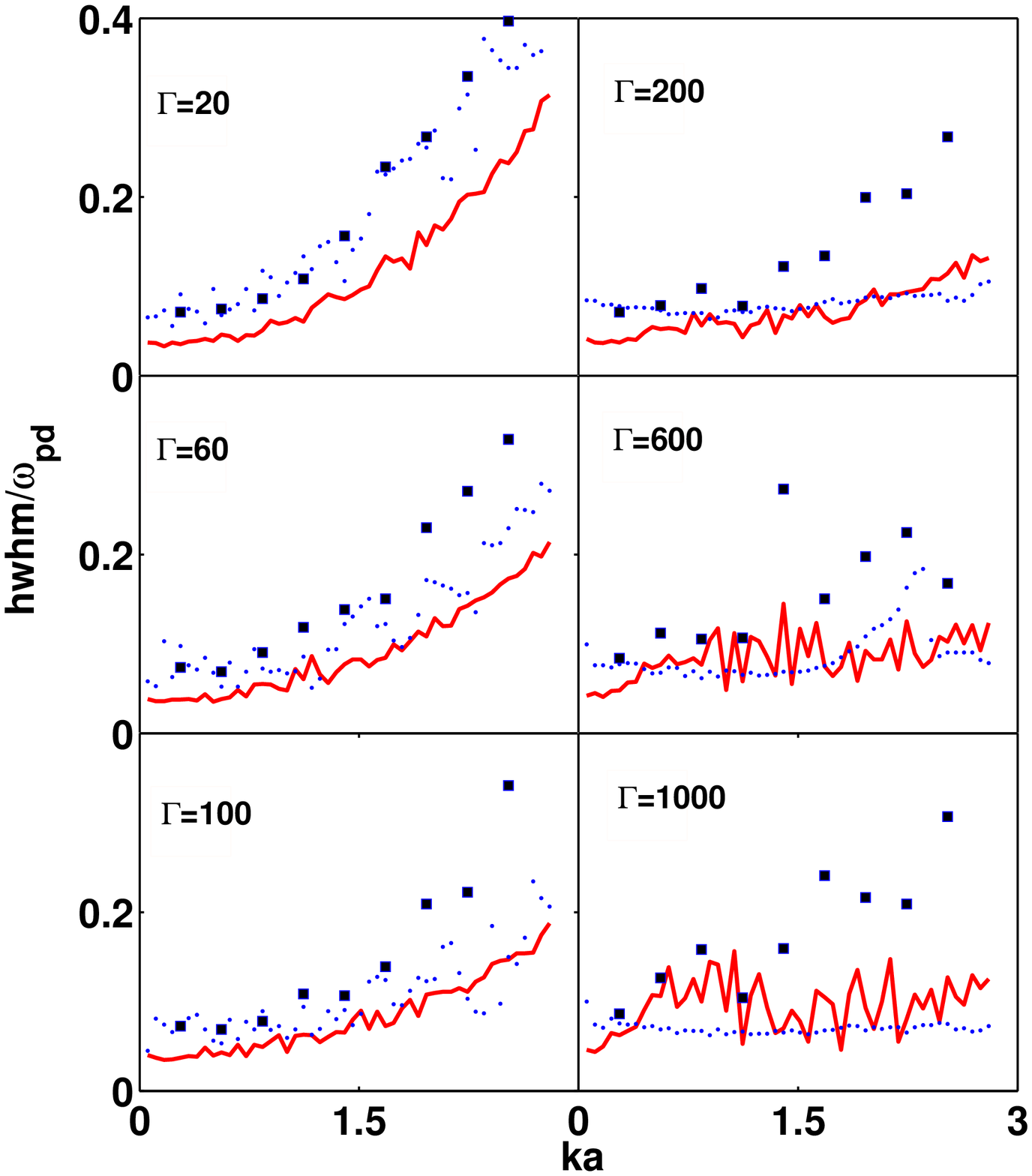}
\caption{\label{hwhm} Half-width at half maximum of the spectral profile curves versus the reduced wave-number $ka$ for
$\kappa=1.0$, $\gamma=0.06$, and $\Gamma=20$, 60, 100, 200, 600, and 1000, with the (blue) dots representing the simulation
data, the (red) noisy solid curve representing the exponential model with finite relaxation time, and the squares representing
the Gaussian model. }
\end{figure}

Finally, in Fig.\ \ref{hwhm}, we show a comparison between the wave-number dependent HWHM values, obtained from the simulation
spectral density profiles and from Eq.\ (\ref{full_spectral_neutral}) where we used both the exponential model with finite
$\tau_l$ values given in Eq.\ (\ref{tau_l}), and the Gaussian model with the frequency moments evaluated from the simulation
spectra via Eq.\ (\ref{second_sum}). One notices that all three sets of data are noisy due to the noise in the simulation
spectra (one recalls that, in the case of the exponential model, the noise stems from $\omega_0(k)$ with $S(k)$ evaluated from
Eq.\ (\ref{structure})). As expected, the HWHM values from the Gaussian model are systematically somewhat larger than those from
the exponential model, but they are both seen to be in reasonably good agreement with the simulation data, even though the
agreement becomes blurred by the increased noise at the highest $\Gamma$ values, say $\Gamma>200$. One also notices in Fig.\
\ref{hwhm} that all damping rates roughly follow a quadratic dependence on $k$, with increasing opening of the parabola as
$\Gamma$ gama increases, and with the limiting value as $k\to 0$ being close to the damping rate due to the neutral drag, given
by $\gamma/2=0.03$.


\subsection{Transverse Wave Mode}

Transverse, or shear waves are expected to only exist for higher coupling strengths $\Gamma$ \cite{Murillo,Nosenko}. In the
following we compare the simulation results with the GH approach for a typical case of $\Gamma=100$.

\begin{figure}
\begin{center}
\includegraphics[width=\textwidth]{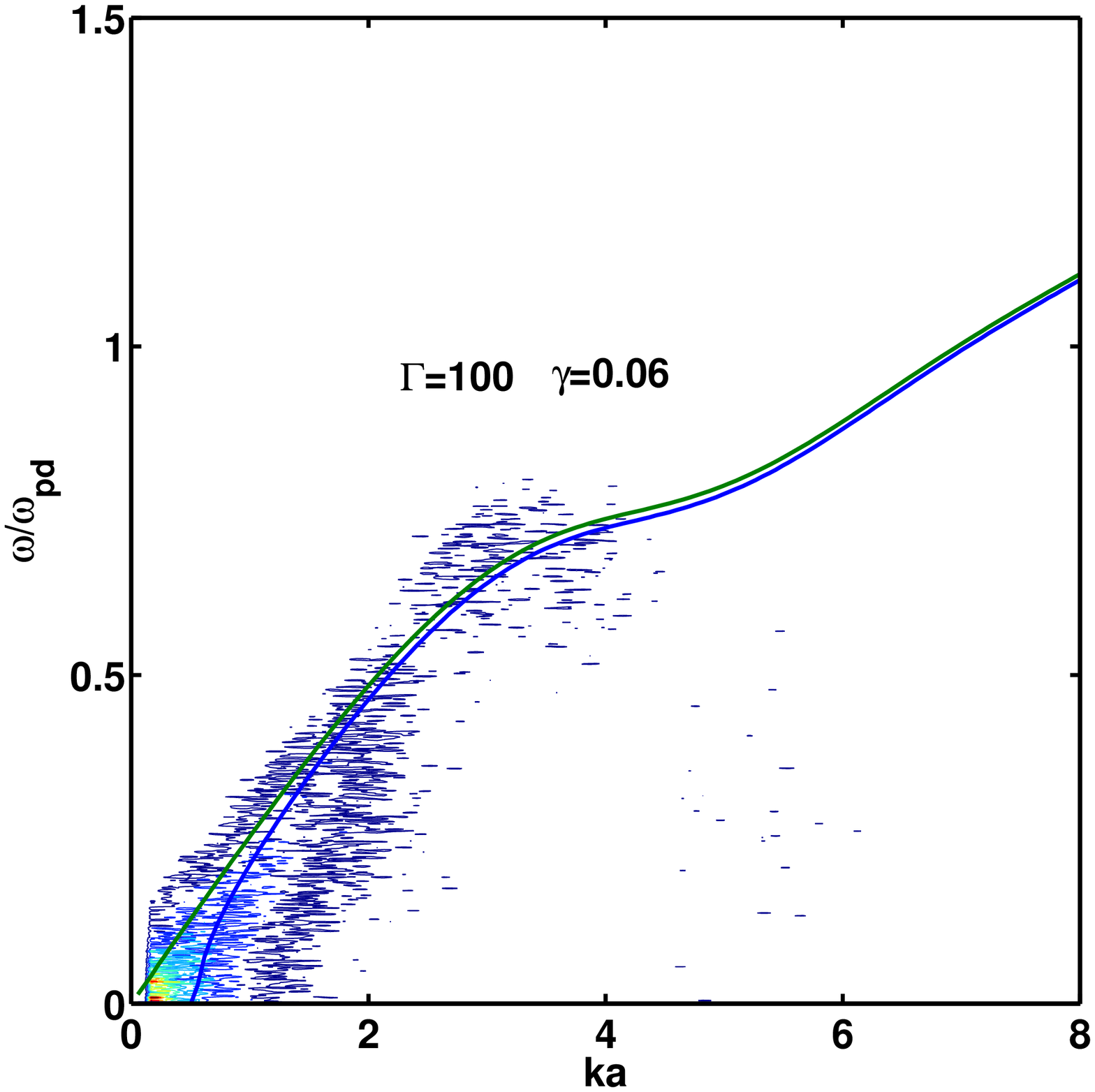}
\caption{\small \label{fig:transWave} Transverse wave dispersion curves against the simulation spectra for $\kappa=1.0$,
$\gamma=0.06$, and $\Gamma=100$, with the upper solid (green) curve and the lower solid (blue) curve representing the
exponential model with an infinitely long and finite relaxation times, respectively.}
\end{center}
\end{figure}

\begin{figure}
\begin{center}
\includegraphics[width=\textwidth]{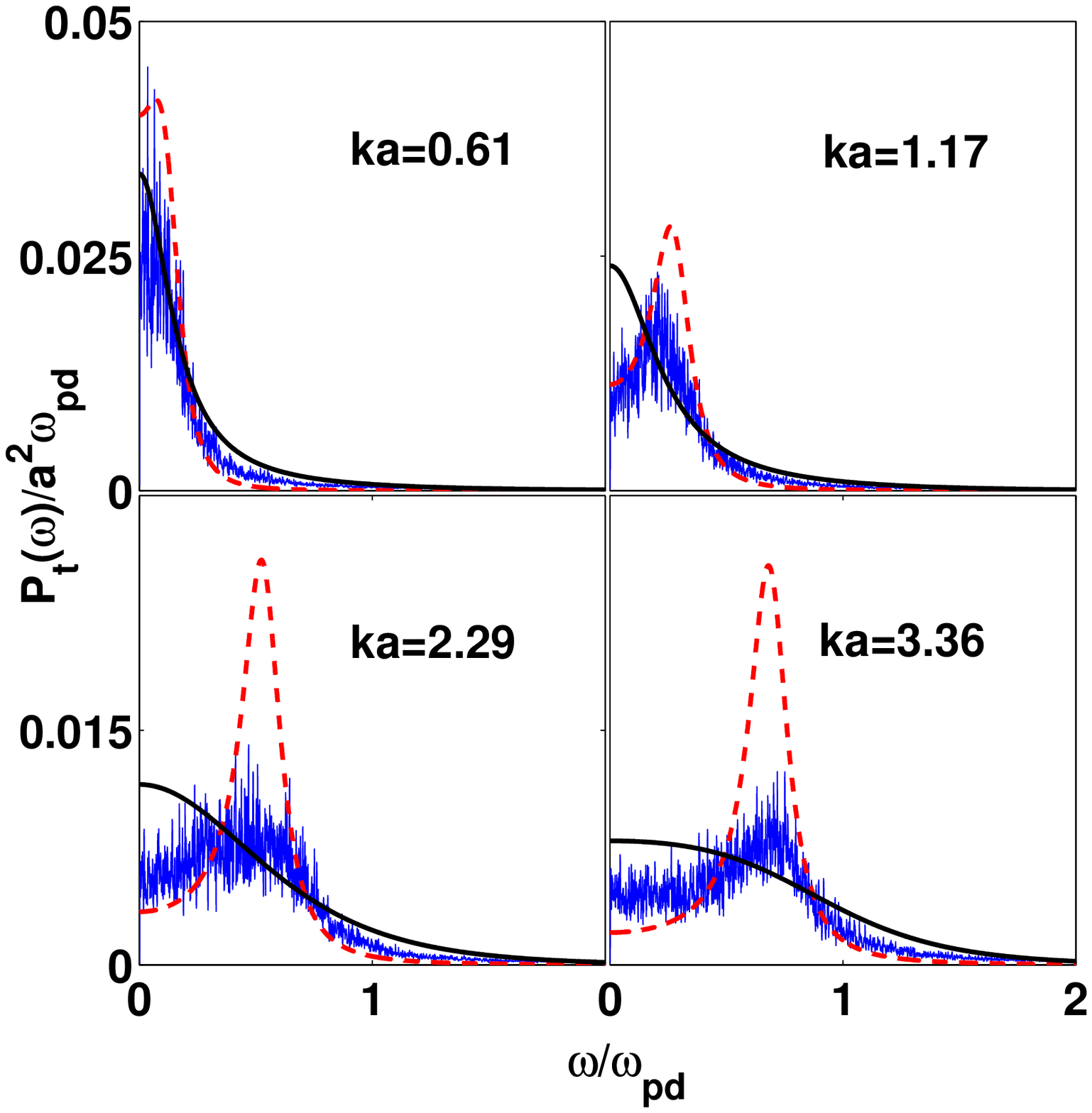}
\caption{\small \label{fig:profile_transWave} Transverse spectral density profile curves versus reduced frequency
$\omega/\omega_{pd}$ for $\kappa=1.0$, $\gamma=0.06$, $\Gamma=100$, and $ka=0.61, 1.17, 2.29$, and 3.36. Simulation data are
shown by the noisy (blue) solid curve, exponential model with finite relaxation time is represented by the dashed (red) curve,
and the Gaussian model is represented by the smooth solid (black) curve. }
\end{center}
\end{figure}

Using Eq.\ (\ref{second_moment_transverse}) to evaluate the second frequency moment from the RDF, we have obtained a dispersion
relation from the peak positions of the spectral density in Eq.\ (\ref{trans_power}) for the exponential model with $\tau_t=5.0$
(note that typical values of $\tau_t$ for the transverse mode are generally found to be larger than those for the longitudinal
mode, in agreement with the remarks of Boon and Yip \cite{Boon}). The results are shown in Fig.\ \ref{fig:transWave}, along with
the dispersion relation in the limit of infinite relaxation time, $\tau_t\to\infty$, given by $\omega=\omega^t_\infty(k)$ with
Eq.\ (\ref{infty_dispersion_transverse}), and are compared with the simulation spectrum for $\Gamma=100$. Both theoretical
dispersion curves show good agreement with the simulation data for $ka< 4$ and, while the simulation spectrum is not conclusive
about the cutoff wave-number, the model with finite $\tau_t$ clearly exhibits a cutoff at $k_c\approx 0.5/a$
\cite{Murillo,Nosenko}.

Further, several profile plots of the spectral density are shown in Fig.\ \ref{fig:profile_transWave} for $\Gamma=100$, where
the results from Eq.\ (\ref{trans_power}) for both the exponential model with $\tau_t=5.0$ and the Gaussian model with the
frequency moments evaluated from the simulation data via Eq.\ (\ref{second_sum}) are compared with the simulation profiles for
$ka=0.61, 1.17, 2.29$, and 3.36. One notices that the agreement of the exponential model with the simulation spectral profiles
is quite satisfactory for the lower two wave-numbers, but the simulation profiles at the two higher wave-numbers are seen to be
much broader than the corresponding peaks from the exponential model. On the other hand, the Gaussian model does not reproduce
the dispersion of the transverse mode at all, since the corresponding profile curves peak at zero frequency for all
wave-numbers, as displayed in Fig.\ \ref{fig:profile_transWave}. Nevertheless, the broadening of the simulation profiles at the
two higher wave-numbers in Fig.\ \ref{fig:profile_transWave} seems to be better echoed by the Gaussian than by the exponential
spectral profile curves.


\section{Concluding Remarks}
We have carried out Brownian Dynamics simulation of a 2D layer of strongly-coupled dusty plasma and extracted the equilibrium
radial distribution function, static structure factor, and the spectral densities for both the longitudinal and transverse dust
acoustic wave modes.

We have then shown that the memory function formalism of the theory of generalized hydrodynamics provides good semi-analytic
results to model these wave modes. In particular, following the approach of Boon and Yip\cite{Boon}, we have developed an
exponential model for the viscosity memory functions that satisfies the second frequency sum rule with the relaxation time being
used as a fitting parameter. With such exponential model we have obtained good fits to the simulation data for the longitudinal
dispersion curves over a wide range of coupling strengths, $20\le\Gamma\le 1000$, as well as reasonable match with the spectral
density profiles for several values of the wave-number $k$. Next, we have extended the theory in a straightforward manner to
satisfy the third and fourth frequency sum rules, providing a parameter-free Gaussian model for the viscosity memory function.
The results using this model provided a successful test for our choice of the relaxation time in the exponential model in
modeling both the dispersion relations and the spectral density profiles.


Moreover, we pointed out that the limits of an infinitely long and infinitesimally short relaxation times in the exponential
model represent two parameter-free models that are fully determined by the equilibrium radial distribution function and the
static structure factor of the system. It was then shown that these two limits provide good upper and lower bounds for the
thermal noise observed in the simulation spectra for the longitudinal mode. Comparisons of the dispersion relations obtained
from these two limits of the exponential model with the dispersion relations obtained from both the Quasi-localized charge
approximation (QLCA) and the extended QLCA (EQLCA) showed that the limit of an infinitely long relaxation time reproduces
remarkably well all the features of the EQLCA model, including the direct thermal effect. On the other hand, the limit of an
infinitesimally short relaxation time was found to reproduce the near-vanishing of the dispersion relation, which is seen in the
simulation spectra at certain short wave-lengths for large coupling strengths, but is not reproduced by the QLCA group of
results.

In addition, we have also tested the memory function formalism within the theory of generalized hydrodynamics against the
simulation results obtained for transverse wave modes at the coupling strength of $\Gamma=100$, and found that the exponential
model with relaxation time used as a fitting parameter can provide a reasonable estimate for the wave dispersion relation,
including the appearance of a cutoff wave-number. Both the peak positions and the widths in the spectral density profiles from
the simulation were well reproduced by the exponential model at long wavelengths, whereas the broadening of these spectra at
short wavelengths was better echoed by the Gaussian model.

The most important finding of the present analysis is that
the wave-number dependent damping rates, which were extracted from the simulation spectra, were found in good agreement with the
results from both the exponential and Gaussian models, testifying to the strengths and potentials of using the memory-function
formalism within the GH theory in describing the damping of the longitudinal dust acoustic wave in strongly-coupled dusty
plasmas.

Finally, we have included the effect of neutral drag due to the collisions of dust particle with neutral molecules in the
background plasma via a (local in time) drag coefficient. While we did not find that the neutral drag affects our modeling of
dispersion relations for both the longitudinal and transverse waves in any significant way, the overall damping rates for the
longitudinal wave showed possibly important roles of the neutral drag at long wavelengths and high coupling strengths where
viscous damping is expected to be reduced. This aspect of the GH theory shall be studied in future work.

In conclusion, the memory function formalism of the generalized hydrodynamics provides a very promising route to analytical
modeling of spectral density of collective modes in strongly coupled 2D dusty plasmas and, in particular it demonstrates a good
account of viscoelastic effects which are not well described by other analytical methods.

\vspace{10mm} This work was supported by the Natural Sciences and Engineering Research Council of Canada. LJH would thank M. S.
Murillo and V. Nosenko for valuable discussions.

\appendix

\section{Frequency sum rules}

Initial-value properties of the current density auto-correlation functions are related to the even-order frequency moments via
\cite{Boon}
\begin{eqnarray}
\langle\omega^{2n}_{l,t}(k)\rangle = (-1)^n\left[\frac{\partial^{2n}}{\partial t^{2n}}C_{l,t}(k,t)\right]_{t=0}, \label{moments}
\end{eqnarray}
whereas all odd-order moments must vanish. We extend here the development outlined by Boon and Yip \cite{Boon} by introducing
the third and fourth frequency sum rules into the GH theory of spectral density. Since these sum rules are used to fix the short
time properties of the model memory functions, we may again initially neglect the neutral drag.

To illustrate this procedure for the longitudinal mode, we differentiate Eq.\ (\ref{new-memory}) twice to obtain the third
derivative as
\begin{eqnarray}
 &&\frac{\partial^3}{\partial t^3}C_l(k,t) = -\frac{k^2v_{th}^2}{S(k)}\frac{\partial}{\partial t}C_l(k,t) - k^2\int _0^tdt'\
C_l(k,t')\frac{\partial^2}{\partial t^2}\phi _l(k,t-t') \nonumber \\
&&- k^2\left[C_l(k,t')\frac{\partial}{\partial t}\phi _l(k,t-t')\right]_{t'=t} - k^2\phi _l(k,0)\frac{\partial}{\partial
t}C_l(k,t),
\end{eqnarray}
which on evaluating at $t=0$, yields
\begin{eqnarray}
\left[\frac{\partial^3}{\partial t^3}C_l(k,t)\right]_{t=0} =  - k^2\left[C_l(k,t')\frac{\partial}{\partial t}
\phi_l(k,t-t')\right]_{t'=t=0}.
\end{eqnarray}
Since the third moment must vanish, we have the condition
\begin{eqnarray}
- k^2C_l(k,0)\left[\frac{\partial}{\partial t}\phi _l(k,t-t')\right]_{t'=t=0}=0,
\end{eqnarray}
showing that the first derivative of the longitudinal viscosity memory function must vanish at initial time. This is to be
expected because a memory function should be viewed as being just another time auto-correlation function with vanishing
derivatives of odd order at the initial time \cite{Boon}. Next, evaluating the fourth moment we find
\begin{eqnarray}
&\frac{\partial^4}{\partial t^4}C_l(k,t) = -\frac{k^2v_{th}^2}{S(k)}\frac{\partial^2}{\partial t^2}C_l(k,t) - k^2\int _0^tdt'\
C_l(k,t')\frac{\partial^3}{\partial t^3}\phi _l(k,t-t') \nonumber\\
&- k^2\left[C_l(k,t')\frac{\partial^2}{\partial t^2}\phi _l(k,t-t')\right]_{t'=t} -k^2\left[\frac{\partial}{\partial
t}C_l(k,t)\frac{\partial}{\partial t}\phi _l(k,t-t')\right]_{t'=t} \nonumber\\
&-k^2\phi _l(k,0)\frac{\partial^2}{\partial t^2}C_l(k,t).
\end{eqnarray}
Again setting $t=0$ we obtain,
\begin{eqnarray}
\left[\frac{\partial^4}{\partial t^4}C_l(k,t)\right]_{t=0} = && -\frac{k^2v_{th}^2}{S(k)}\left[\frac{\partial^2}{\partial
t^2}C_l(k,t)\right]_{t=0} \nonumber\\
&&- k^2C_l(k,0)\left[\frac{\partial^2}{\partial t^2}\phi _l(k,t-t')\right]_{t'=t=0}\nonumber\\
&&-k^2\phi _l(k,0)\left[\frac{\partial^2}{\partial t^2}C_l(k,t)\right]_{t=0}.
\end{eqnarray}
Substituting $\phi_l(k,0)$ from Eq.\ (\ref{initialvalue-memory}) and invoking Eq.\ (\ref{moments}) with $n=1$ and 2,
we obtain Eq.\ (\ref{memory_2nd_initial}) in the main text.

\clearpage

\section*{References}

\end{document}